\documentclass{article}

\usepackage{arxiv}

\usepackage[utf8]{inputenc} % allow utf-8 input
\usepackage[T1]{fontenc}    % use 8-bit T1 fonts
\usepackage{hyperref}       % hyperlinks
\usepackage{url}            % simple URL typesetting
\usepackage{booktabs}       % professional-quality tables
\usepackage{amsfonts}       % blackboard math symbols
\usepackage{nicefrac}       % compact symbols for 1/2, etc.
\usepackage{microtype}      % microtypography
\usepackage{lipsum}		% Can be removed after putting your text content
\usepackage{graphicx}
\usepackage{doi}
\usepackage{soul}
\usepackage{color}
\usepackage{siunitx}
\usepackage{xspace}
\usepackage{changes}
\usepackage{xcolor}

\newcommand{\MR}{MRR\xspace}
\newcommand{\MRs}{MRRs\xspace}
\newcommand{\SP}{SP\xspace}

\title{Reservoir computing with all-optical non-fading memory in a self-pulsing microresonator network}

\author{ 
    \href{https://orcid.org/0000-0002-6587-2614}{\includegraphics[scale=0.06]{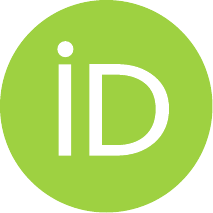}\hspace{1mm}Alessio Lugnan\(^1\),} 
\And
	\href{https://orcid.org/0000-0002-3361-133X}{\includegraphics[scale=0.06]{orcid.pdf}\hspace{1mm}Stefano Biasi\(^1\),} 
\And
    \href{https://orcid.org/0009-0001-3195-8620}{\includegraphics[scale=0.06]{orcid.pdf}\hspace{1mm} Alessandro Foradori\(^{2,1}\),}
\And
    \href{https://orcid.org/0000-0001-6259-464X}{\includegraphics[scale=0.06]{orcid.pdf}\hspace{1mm} Peter Bienstman\(^2\),}
\And
	\href{https://orcid.org/0000-0001-7316-6034}{\includegraphics[scale=0.06]{orcid.pdf}\hspace{1mm}Lorenzo Pavesi\(^1\)}
  	 \\ \\
    \(^1\) Nanoscience Laboratory, Department of Physics, University of Trento, 38123 Trento, Italy
    \\
    \(^2\) Photonics Research Group, Department of Information Technology, Ghent University-imec, Ghent, Belgium
    \\
    \\
    \\
    Corresponding author: alessio.lugnan.1@unitn.it
    \\
}

\renewcommand{\headeright}{Research article}
\renewcommand{\undertitle}{Research article}
\renewcommand{\shorttitle}{}

\hypersetup{
pdftitle={Reservoir computing with all-optical non-fading memory in a self-pulsing microresonator network},
pdfsubject={q-bio.NC, q-bio.QM},
pdfauthor={},
pdfkeywords={},
}

\begin{document}
\maketitle

\begin{abstract}
Photonic neuromorphic computing may offer promising applications for a broad range of photonic sensors, including optical fiber sensors, to enhance their functionality while avoiding loss of information, energy consumption, and latency due to optical-electrical conversion. However, time-dependent sensor signals usually exhibit much slower timescales than photonic processors, which also generally lack energy-efficient long-term memory. To address this, we experimentally demonstrate a first implementation of physical reservoir computing with non-fading memory for multi-timescale signal processing. This is based on a fully passive network of 64 coupled silicon microring resonators. Our compact photonic reservoir is capable of hosting energy-efficient nonlinear dynamics and multistability. It can process and retain input signal information for an extended duration, at least tens of microseconds. Our reservoir computing system can learn to infer the timing of a single input pulse and the spike rate of an input spike train, even after a relatively long period following the end of the input excitation. We demonstrate this operation at two different timescales, with approximately a factor of 5 difference. This work presents a novel approach to extending the memory of photonic reservoir computing and its timescale of application.
\end{abstract}

% keywords can be removed

\section{Introduction}
\label{sec:intro}
Photonics is an attractive platform for neuromorphic computing, offering high throughput, low latency, and energy-efficient linear operations \cite{shastri2021photonics, pavanello2023special, xu2023reconfigurable}.  Indeed, photonic hardware can execute the linear operations behind inter-layer connections in artificial neural networks (ANNs), which is particularly useful for accelerating large ANN models in data centers. In contrast, edge computing applications typically require smaller, more specialized ANNs that operate on compact and affordable hardware, capable of continuous learning or easy retraining for new conditions \cite{christensen20222022, covi2021adaptive}.  In this context, photonic neuromorphic hardware is particularly attractive if directly applied to information encoded in the optical domain, such as telecom data through optical fibers and optical sensors output. This would enable avoiding or reducing latency and power consumption due to optical-to-electrical conversion.  However, competitive ANN solutions require scalable and energy-efficient nonlinear operations and memory (the latter being necessary for processing time-dependent signals). These requirements remain difficult to achieve in photonics due to the lack of direct photon-photon interaction.  

As we demonstrate in this paper, these key properties can be found in silicon microring resonators (\MRs), which are simple CMOS-compatible devices \cite{bogaerts2012silicon} widely used in photonics for a variety of applications, from wavelength filtering to sensing.
\MRs provide an exceptional platform for light-matter interactions, significantly enhancing nonlinear effects. At 1550 nm, these nonlinearities are triggered by two-photon absorption (TPA) and the associated free carrier absorption and dispersion (FCA and FCD) \cite{leuthold2010nonlinear, borghi2017nonlinear}. TPA generates free carriers in the ring waveguide. Subsequently, these carriers absorb light and undergo thermalization, leading to heat production. This process raises the silicon's temperature, altering its refractive index through the thermo-optic (TO) effect. FCD and TO effects cause blue and red shifts, respectively, in the resonance frequency of the \MR \cite{johnson2006selfinduced,biasi2022effect}. Additionally, FCD and TO effects have distinct relaxation times and dependencies on the optical field amplitude. Specifically, the thermal and carrier lifetimes are generally around 60-280 ns and 1-45 ns, respectively \cite{vaer2012simplified, borghi2021modeling}. Due to these differing characteristics, the combination of TO, TPA, and FCD effects can result in self-pulsing (\SP) oscillations \cite{priem2005optical, pavesi2021thirty}. Here, a simple continuous wave input signal, with a wavelength near the \MR's resonance wavelength, is transformed into an oscillating output due to the nonlinear and oscillatory resonance frequency shift \cite{priem2005optical, pavesi2021thirty}. Notably, by arranging multiple silicon \MRs in series, a diverse range of dynamic \SP responses, including chaotic ones, can be achieved by varying the input laser power and wavelength \cite{mancinelli2014chaotic}.

Building on these effects, prior studies have demonstrated the potential of silicon \MRs as energy-efficient artificial spiking neurons with a minimal chip footprint \cite{van2012cascadable, xiang2020all, xiang2022all, lugnan2022rigorous, zhang2024chip, biasi2024taiji}. Furthermore, silicon \MRs have been extensively utilized in the broader field of neuromorphic computing \cite{biasi2024photonic}. In recent experiments, we have shown that it is feasible to directly connect multiple \MRs together, creating relatively large and scalable neural networks on a photonic chip \cite{biasi2023array, lugnan2023emergent}. Moreover, we have recently demonstrated that small networks of up to 3 coupled silicon \MRs can host complex \SP dynamics and multi-stability, enabling the storage of input information for much longer durations (at least \SI{10}{\micro\second}) compared to what silicon nonlinear effects intrinsically allow \cite{biasi2024exploring}. In this work, we exploit this long-term memory effect in a larger network of 64 coupled silicon MRRs for hardware-based machine learning (ML). Input information, specifically single pulse timing and pulse train rate, is retrieved by a linear regressor or classifier by reading the photonic network output long after it has been fed with the input signal. 

This ML approach can be considered as a special case of hardware-based reservoir computing (RC) \cite{jaeger2001echo}. However, in its original definition RC works with fading memory, while for the tasks we consider in this paper, we exploit the non-fading memory of our \MRs network. The RC method essentially involves the use of a recurrent neural network (or a non-linear dynamical system in general), called a \textit{ reservoir}. The parameters of this reservoir are not trained; instead, they are kept fixed after a random initialization \cite{cucchi2022hands}. The only trainable part is a readout linear regressor or classifier, which is applied to some of the reservoir neurons (or states in general). The function of the reservoir is to provide memory and expand the dimensionality of the input signal, greatly enhancing the computational power of the trainable linear readout. Although it may not be as computationally powerful as fully trained recurrent neural networks or transformers, RC is particularly attractive due to its simple and fast training process. Most importantly, in the context of this work, it is hardware-friendly. In fact, any sufficiently complex dynamical physical system, such as a bucket of water \cite{fernando2003pattern}, can in principle be used as a reservoir, since there is no need to control or fully observe its internal dynamics for training. This advantage makes hardware-based RC a popular neuromorphic computing approach \cite{tanaka2019recent, lugnan2020photonic}, which is especially suitable for edge computing.

\begin{figure*}[t!]
	\centering
	\includegraphics[width=0.8\textwidth]{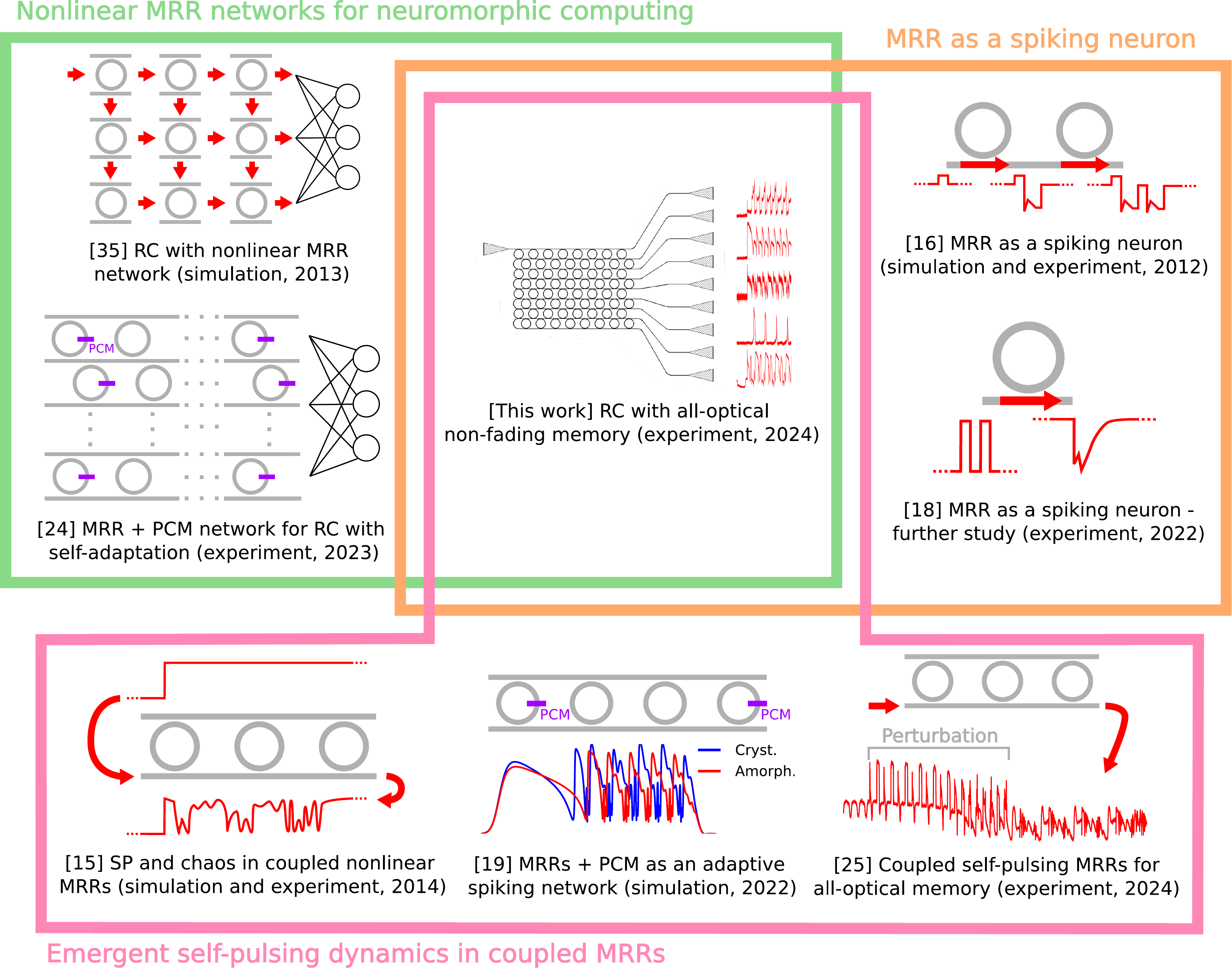}
	\caption{\textbf{Relevant previous works.} Here we summarize previous works that inspired this research, with a short description and a sketch about the most relevant content. We divide these works into three groups, corresponding to three ways of considering nonlinear MRRs for neuromorphic computing.}
 \label{fig:0_relWorks}
 \end{figure*}
 
Inspired by the works summarized in Fig. \ref{fig:0_relWorks}, in this paper, we experimentally demonstrate a fully passive all-optical reservoir with non-fading memory. This reservoir can be read out after a relatively long period (at least tens of microseconds) following its input excitation to perform signal processing and detection. This capability is based on the sensitive yet stable dynamical states and multistability hosted by our photonic network, which can be altered by an input signal, allowing the information to be stored. Importantly, this long term memory effect can be used to match the effective timescale of RC operations with the input signal timescale, which is in general a critical and complex challenge of non-digital neuromorphic computing \cite{jaeger2021dimensions} and of physical RC in particular \cite{yan2024emerging}. For example, widespread optical sensing applications such as fiber sensing \cite{lu2019distributed}, typically work within much slower timescales (milliseconds or slower) than those of photonic networks (microseconds or faster). Our work provides a novel path to bridge this gap and achieve energy efficient all-optical preprocessing for photonic smart sensing applications. Further advantages of our integrated photonic reservoir are: small footprint (0.15 mm\(^2\)), relatively low power consumption (from few to few tens of mW), no need of external connections or wiring as it can be reconfigured by changing the excitation wavelength or power at its input, simple fabrication, multi-wavelength capability enabling network dimension expansion and/or parallel computing. Moreover, we previously showed that similar \MR-based networks can be employed to tackle relatively complex ML tasks, such as handwritten digits classification \cite{lugnan2023emergent, lugnan2024large}. Finally, it should be stressed that our implementation was inspired by a simulation work published a decade ago \cite{mesaritakis2013micro}, which first proposed using a matrix of coupled \MRs driven in the nonlinear regime for RC.

The rest of the article is organized as follows. In the Results section, we first present the principle of operation for the proposed photonic RC with non-fading memory (Section \ref{subsec:principle}). Secondly, we describe the input waveforms used for the measurements and the considered ML tasks (Fig. \ref{subsec:MLtasks}). Then, we discuss the obtained ML performances (\ref{subsec:MLperf}). Following that, we present our conclusions. Finally, in the Methods section, we describe further technical details.

\section{Results}
\label{sec:Results}
\subsection{Principle of operation}
\label{subsec:principle}
\begin{figure*}[t!]
	\centering
	\includegraphics[width=0.8\textwidth]{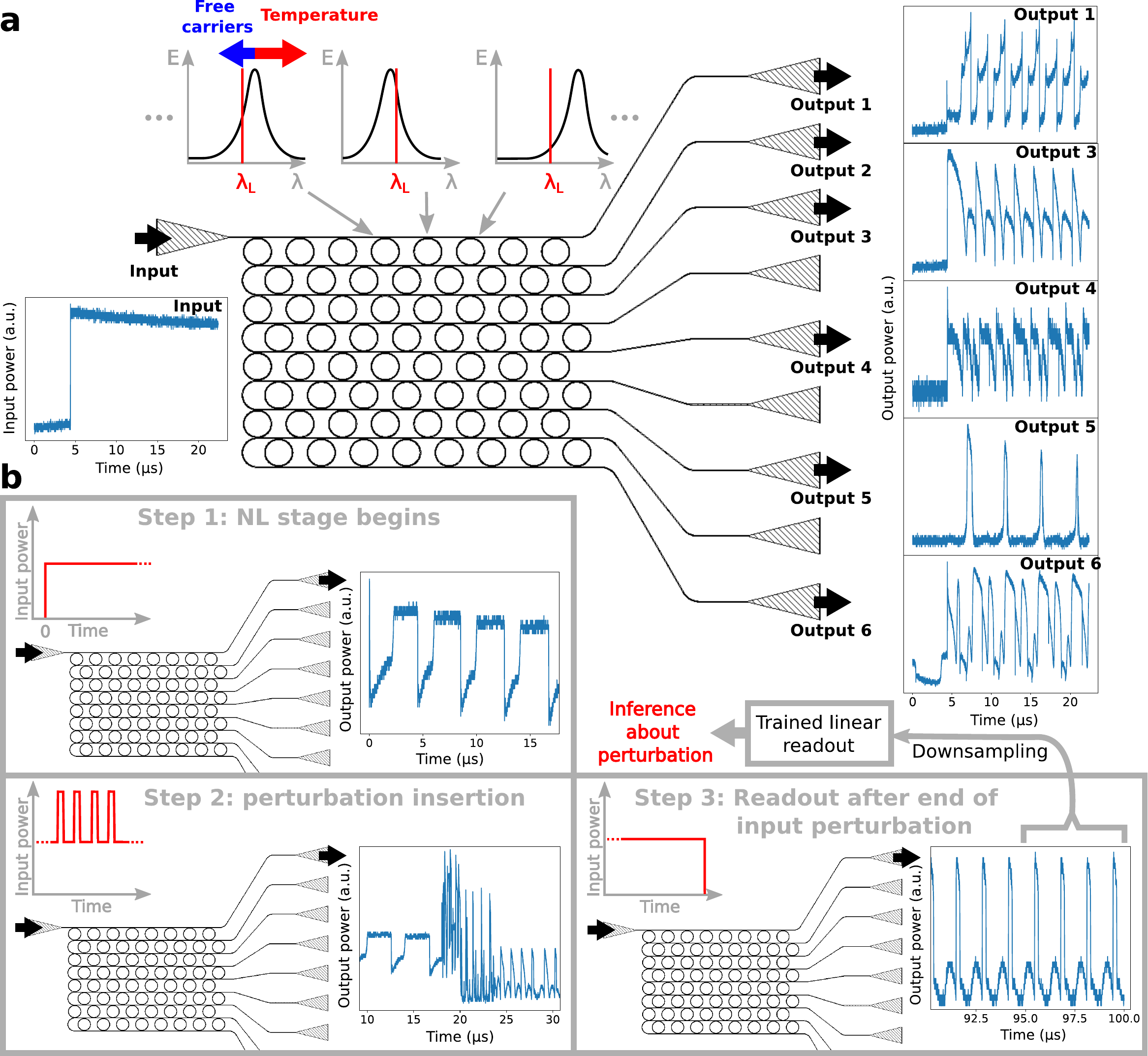}
	\caption{\textbf{Working principles of photonic RC with non-fading memory.} \textbf{a} Our reservoir  (schematics at the center) consists of a network of coupled silicon \MRs, whose resonance is centered at different wavelengths due to fabrication errors (see plots on the top, showing \MR stored energy v.s. wavelength for few \MRs). Moreover, if the input laser power is high enough (milliwatts), the resonance wavelength of the excited \MRs is shifted because of silicon nonlinear effects (due to the free carriers and temperature) and can give rise to multistability and complex network dynamics due to the coupled dynamics of multiple \MRs. The reservoir is fed by an input signal (example on the left, blue line), which is coupled to the input grating (triangle), and produces different outputs (example on the right, blue lines), which are collected by the various output ports via output gratings. \textbf{b} Measurement steps. At first, Step 1, the input power (left, red line) is increased to drive the network into a nonlinear regime, e.g. a self-pulsing (\SP) response (right, blue line). Then, Step 2, a perturbation (a single pulse or a pulse train) is added to the constant input. The perturbation can alter the network dynamics in a non-fading way, as shown in the example output lineshape on the right. Finally, Step 3, a linear readout is trained to infer some perturbation properties from the network response long after the perturbation has ended. The input power, Step 4 - not shown, is then lowered to reset the network state.}
 \label{fig:1_network}
 \end{figure*}

Our neuromorphic hardware consists of a 8 × 8 matrix of silicon MRRs, with a relatively low quality (Q) factor of about 6500. The 64 \MRs are coupled in an add-drop configuration by straight silicon waveguides. These Q factors, while providing a low field enhancement for nonlinear operations, allow coupling statistically more MRRs for a given input laser wavelength. In fact, a low Q factor implies a broad \MR resonance, more light that is transmitted or dropped and a low sensitivity of the resonance wavelengths to fabrication defects. We employed a grating coupler as the input port (shown on the left of Fig. \ref{fig:1_network} a) and used 6 grating couplers as output ports (shown on the right of Fig. \ref{fig:1_network} a). The input signal consists of a modulated coherent light (wavelength around \SI{1550}{\nano\meter}) to generate a time-dependent input signal.  Each \MR is resonant in a small wavelength interval (full width at half maximum, FWHM$\simeq$0.24 nm). When not excited, each \MR resonance is centered at different wavelengths, as they are randomly displaced due to unavoidable fabrication inaccuracies (see examples at the top of Fig. \ref{fig:1_network} a, where the stored energy vs. wavelength is plotted for few \MRs). This means that each fabricated network is unique and that it is not possible to predict a \MR resonance wavelength before fabrication. However, once the \MRs are manufactured, their resonances are stable in the absence of strong optical excitation (linear regime) and if the chip temperature remains relatively constant (we applied a thermostat to the chip holder limiting fluctuations to less than 0.1 °C). For intuitive insight, we can approximate the considered \MR response as follows: if the light incoming through a straight waveguide has a wavelength that is fully resonant, it is mostly redirected by the ring to the other coupled straight waveguide, towards the opposite direction w.r.t. the input. On the other hand, if the incoming light is not resonant, it mostly travels through the straight waveguide as if no \MR was there; in in-between cases, the input light is split among the two output directions, with proportions depending on how close the input light is to resonance. For more in-depth details about the \MR theory, we refer the reader to the relevant literature \cite{bogaerts2012silicon, biasi2024photonic}. 

High enough input power levels (typically on the order of milliwatts or higher) drive a resonant \MR into the nonlinear regime (see Fig. \ref{fig:1_network} a, top inset). In this case, the \MR resonance is not fixed anymore but shifted in wavelength by the nonlinear silicon effects. In our network, a sufficiently powerful constant input can induce nonlinear dynamics involving several rings, causing complex time-dependent responses at the output ports (see exemplary input and output waveforms in Fig. \ref{fig:1_network} a). Importantly, the output signals are strongly dependent on the input power, wavelength, and output port, allowing the generation of a great variety of dynamical responses. These can be explored to find the most suitable ones for a target computational task. Indeed, this is the approach we employ in this work.

In particular, we focus on a specific way to excite and use the network response (here referred to as \textit{measurement}), to demonstrate a photonic reservoir with long-term and non-fading memory. Each performed \textit{measurement} can be divided in three consecutive steps (with reference to Fig. \ref{fig:1_network} b):
\begin{enumerate}
    \item[Step 1:] \textit{Start of nonlinear (NL) stage.} The input power (red line in Fig. \ref{fig:1_network} b) is quickly increased from a low level (linear regime) to a constant high level to drive the network in the nonlinear regime in the network. The role of this stage is to sustain the nonlinear regime and/or \SP dynamics in the photonic network.  
    \item[Step 2:]  \textit{Perturbation insertion.} A time-dependent perturbation (either a single pulse or a pulse train), whose duration is much shorter than the NL stage duration, is added to the constant input optical power of the ongoing NL stage.
    \item[Step 3:] \textit{Readout after end of input perturbation.} Well after the input perturbation has ended and right before the end of the NL stage, the output signal is sampled (we name the sampling time interval the \textit{readout interval}). These values are fed into the ML linear readout, which is trained to infer the perturbation information from the network response. 
    \item[Step 4:] \textit{End of NL stage.} After the readout, the NL stage ends with a quick drop of the input power. This leaves the network in the linear regime for a long enough time to reset its memory (given by the free carriers and the temperature variation in the \MRs).
\end{enumerate}
As we explain later in more details, in this work we consider two different perturbation types (single pulse and pulse train), two different NL stages and perturbation timescales, and two different types of ML methods (linear regression and classification using logistic regression). For each of these cases, we tackle two different problems: inference on the timing of a single pulse and inference on the spike rate of a train of pulses. In this paper, each combination of these elements is referred to as a \textit{ML task}.

\begin{figure*}[t!]
	\centering
	\includegraphics[width=1\textwidth]{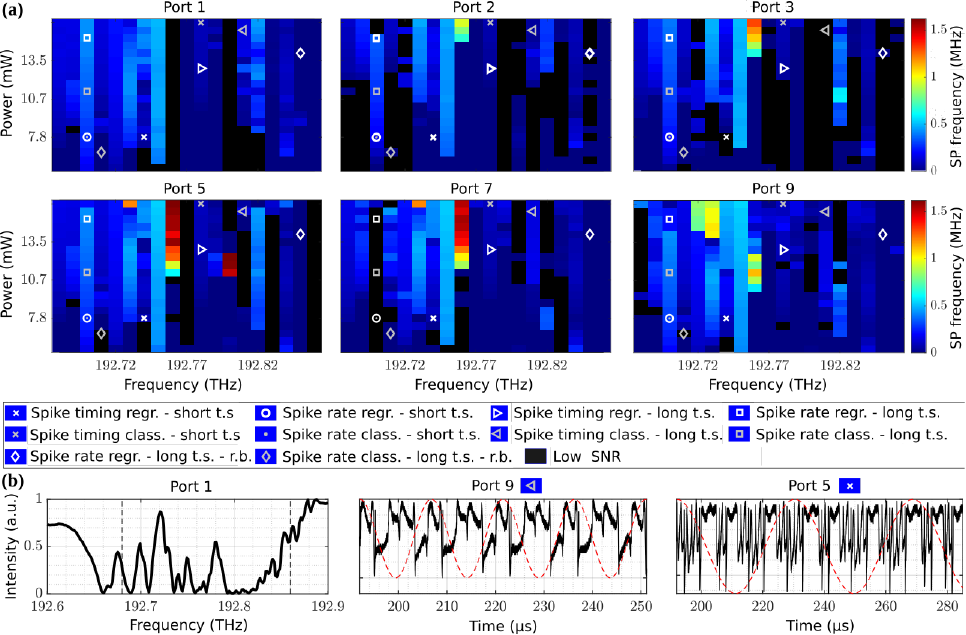}
	\caption{\textbf{Self-pulsing frequency as a function of laser frequency and power.} \textbf{a} Color-maps of the \SP frequency (i.e. inverse of the fundamental period of the \SP waveform) at the different  output ports. Ten different markers on the maps (see the legends below the maps) indicate the input parameter combinations that optimize the performance of a specific ML task (described in Sections \ref{subsec:MLtasks} and \ref{subsec:MLperf}). In the legend, `t.s.' and `r.b.' respectively stand for `timescale' and `random beginning' of perturbation. \textbf{b} The left plot shows the transmission spectrum of the reservoir in the linear regime measured at output power 1, with dotted vertical lines indicating the spectral window where the \SP frequency maps were measured. The middle and right plots display two examples of \SP output waveforms measured at port 9 and port 5, respectively, for the set of parameters represented by the symbol on the top of the panel. A fit with a sinusoidal lineshape (dotted red line) allows to estimate the fundamental \SP frequency reported in the color-maps.}
 \label{fig:2_SPmap}
 \end{figure*}

Importantly, for each \textit{ML task}, we repeated a corresponding \textit{measurement} for:
\begin{itemize}
    \item 19 different input optical frequencies that are resonant with some \MRs in the network. We used frequencies from \SI{192.68}{\tera\hertz} to \SI{192.86}{\tera\hertz}, with steps of \SI{0.01}{\tera\hertz}. Moreover, an additional non-resonant frequency (namely \SI{192.60}{\tera\hertz}) was employed as a reference. It corresponds to the case where the input signal is directly transmitted to the output without resonating in any \MRs. We used this reference measurement as a baseline to estimate the performance improvement due to the reservoir, that is, the \MRs array.
    \item 20 different input optical power levels, namely with an estimated average on-chip power from $\simeq$ \SI{5.5}{\milli\watt} to $\simeq$ \SI{16.4}{\milli\watt}, with steps of $\simeq$ \SI{0.57}{\milli\watt}. For the baseline, we used only the maximum power level, in order to maximize the signal-to-noise ratio.
\end{itemize}
Therefore, we performed a total of 380 measurements per task, plus one reference measurement with a non-resonant laser frequency. It should be stressed that, in most of these measurements, the NL stage drove the network into a dynamical state due to the coupled \SP of the involved \MRs. This can be observed in Fig. \ref{fig:2_SPmap} a, where 2D color maps show the estimated \SP output frequencies for the different output ports. We observe a diverse range of \SP responses, with \SP frequencies ranging from about \SI{9.4}{\kilo\hertz} to \SI{1.6}{\mega\hertz}. These maps provide valuable insights into the timescales of the network dynamics. Moreover, in Fig. \ref{fig:2_SPmap} a we marked the combinations of laser frequency and power that optimize the performances of the ML tasks. These are discussed in the next sections. Looking at the \SP maps, we can already see that different combinations of input parameters optimize different tasks, demonstrating the versatility of our RC implementation.

Finally, in Fig. \ref{fig:2_SPmap} b left plot, we show a transmission spectrum of our photonic reservoir in the linear regime, measured at output port 1. It can be seen that the single \MRs resonances overlap in a large structured resonance band, roughly from \SI{192.62}{\tera\hertz} to \SI{192.88}{\tera\hertz}. The detailed analysis of the various contributions to this transmission lineshape is beyond the scope of this work \cite{mancinelli2014chaotic}. The dotted vertical lines show the laser frequency range employed for our measurements. The out-of-resonance baseline measurement has an input frequency located at the beginning of the drawn spectrum. Two examples of the output transmission waveforms due to the complex \SP dynamics are shown on the right in Fig. \ref{fig:2_SPmap} b. A fit of these waveforms with a sinusoidal waveform (dotted lines in Fig. \ref{fig:2_SPmap} b) yields the \SP frequencies reported in the \SP maps. It should be noted that the \SP dynamics can be significantly faster than the \SP frequency.
% 10 KHz port 2 to 1.62 MHz port 5
%<<CHeck with Stefano the different scales on the colorbars! And fill in SP frequency range in text!>>

\subsection{Input waveforms and ML tasks}
\label{subsec:MLtasks}
\begin{figure*}[t!]
	\centering
	\includegraphics[width=0.8\textwidth]{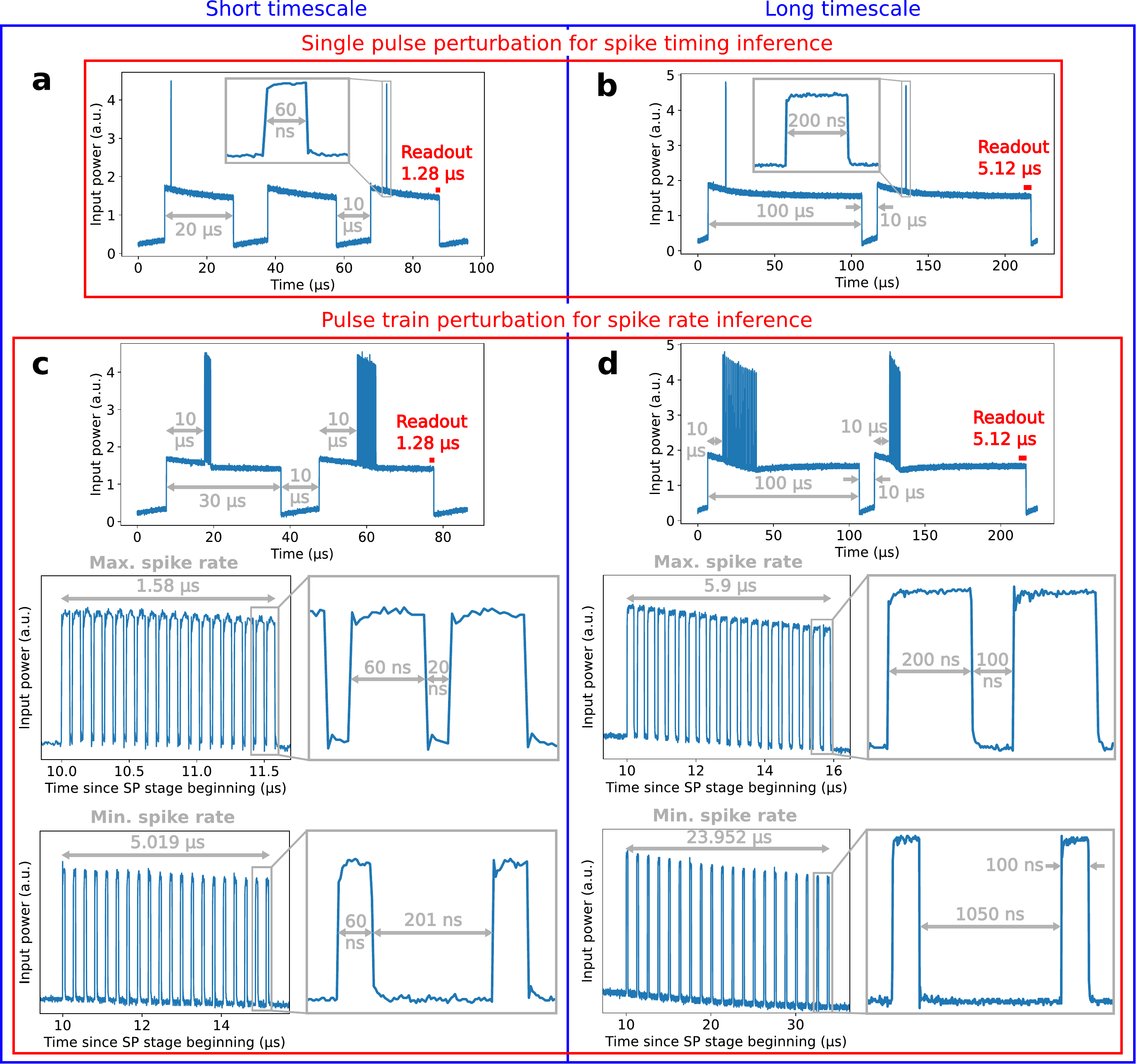}
\caption{ \textbf{Input waveform types, corresponding to different ML tasks.} \textbf{a} Example input waveform segment where the NL stages are perturbed by a single pulse at different timings w.r.t. the NL stage start. (The unperturbed NL stage in the middle was employed as reference and can be ignored.) We marked with a red horizontal bar the short NL stage interval whose corresponding network response is fed into the trainable linear readout. \textbf{b} Similar to \textit{a}, but with longer timescale. \textbf{c} Example input waveform segment (top) where the NL stages are perturbed by a sequence of 20 pulses, always starting at the same time w.r.t. the NL stage start, but with different pulse rates. The plots below  respectively show the detail of the perturbation with highest and lowest spike rates. \textbf{d} Similar to \textit{c}, but with longer timescale.}
	\label{fig:3_tasks}
\end{figure*}

In this work, we mainly performed 5 measurement sessions (each comprising the aforementioned 381 measurements with varying input frequency and power) distinguished by different input optical waveforms employed to excite our photonic network (see Fig. \ref{fig:3_tasks}), each being a sequence of NL stages (see Section \ref{subsec:methods_ML} for further details):
\begin{enumerate}
    \item Single-spike perturbation for \textit{spike timing} inference: the NL stages are perturbed by a single pulse. This can have 50 different timings w.r.t. the NL stage beginning.
    \begin{enumerate}
        \item \textit{Short timescale} version (Fig. \ref{fig:3_tasks} a): the NL stage is \SI{20}{\micro\second} long and the perturbing pulse duration is \SI{60}{\nano\second}. 50 spike timings from 0 to \SI{9.8}{\micro\second}, with a time step of \SI{0.2}{\micro\second}, were used.
        \item \textit{Long timescale} version (Fig. \ref{fig:3_tasks} b): the NL stage is \SI{100}{\micro\second} long and the perturbing pulse duration is \SI{200}{\nano\second}. The 50 spike timings range from \SI{10}{\micro\second} to \SI{19.8}{\micro\second}, with time steps of \SI{0.2}{\micro\second}.
    \end{enumerate}
    \item Spike train perturbation for \textit{spike rate} inference: the NL stage is perturbed by a sequence of 20 equidistant pulses. We consider 20 different spike rate values.
    \begin{enumerate}
        \item \textit{Short timescale} version (Fig. \ref{fig:3_tasks} c): the pulse duration is of \SI{60}{\nano\second}. In the \textit{maximum} spike rate case, the pulses are separated by \SI{20}{\nano\second} and the pulse train is \SI{1.58}{\micro\second} long, starting always \SI{10}{\micro\second} after the start of the NL stage. In the \textit{minimum} spike rate case, the pulses are separated by \SI{201}{\nano\second} and the pulse train is \SI{5.019}{\micro\second} long.
        \item \textit{Long timescale} version (Fig. \ref{fig:3_tasks} d): the pulse duration is \SI{200}{\nano\second}. In the \textit{maximum} spike rate case, the pulses are separated by \SI{100}{\nano\second} and the pulse train is \SI{5.9}{\micro\second} long, starting always \SI{10}{\micro\second} after the start of the NL stage. In the \textit{minimum} spike rate case, the pulses are separated by \SI{1050}{\nano\second} and the pulse train is \SI{23.952}{\micro\second} long.
        \item Long timescale with \textit{random perturbation start}: in contrast with the previous two cases, in this measurement session the pulse train starts at random times w.r.t. the start of the NL stage, ranging from \SI{10}{\micro\second} to \SI{30}{\micro\second}. The other parameters are equal to the long-timescale case.
    \end{enumerate}
\end{enumerate}

A ML dataset is created for each measurement (thus obtaining 381 datasets per measurement session), by extracting a small final part of the NL stages from the network response at each output port. In particular, for the short and long timescale cases, we considered of each NL stage the last \SI{1.28}{\micro\second} (400 oscilloscope samples) and the last \SI{5.12}{\micro\second} (1600 oscilloscope samples) respectively. Note that these traces are downsampled before they are fed to the linear readout (the best among 4 downsampling ratios is chosen). In addition, to generate a ML dataset that better covers the variations due to measurement noise and random parameters, each measurement comprises different repetitions of NL stages with the same nominal parameter values. More details are provided in the Methods, Section \ref{subsec:methods_ML}. 

Each obtained ML dataset is employed to train the following ML linear readout models, for the following ML tasks (all with optimized L2 regularization): 
\begin{enumerate}
    \item \textit{Linear regression}.
    \begin{enumerate}
        \item For \textit{spike timing regression}: the linear readout is trained to return the timing of the single perturbing pulse.
        \item For \textit{spike rate regression}: the linear readout is trained to return the rate of the perturbing pulse train.
    \end{enumerate}
    \item \textit{Classification} (2-classes, logistic regression).
    \begin{enumerate}
        \item For \textit{spike timing classification}: the readout is trained to distinguish between perturbations timings before and after a certain timing threshold. In order to make the evaluation of this task more general, we repeated training and testing using 4 different timing thresholds, roughly situated at 1/5, 2/5, 3/5 and 4/5 of the employed timing values, and we averaged the obtained performances. 
        \item For \textit{spike rate classification}: the readout is trained to distinguish between spike rates below and above a certain rate threshold. As in the previous case, we repeated training and testing using 4 different rate thresholds, roughly situated at 1/5, 2/5, 3/5 and 4/5 of the employed rate values.
    \end{enumerate}
\end{enumerate}
We consider linear regression problems to be more challenging than classification problems. Linear regression aims to associate different continuous values with each instance, rather than simply grouping them into two classes, while using the same number of trainable readout weights. We believe that linear regression provides general insights and elements on the suitability of our reservoir for solving the tasks at hand. In contrast, classification performance (accuracy, in this work) offers a more practical and quantitative measure of how well our RC system groups input perturbations based on their relevant properties.

In order to provide further insight on the practical applicability of the proposed neuromorphic computing framework, we investigate different \textit{ML task variations}. For the \textit{linear regression} problem, we evaluate two variations:
\begin{enumerate}
    \item \textit{Seen} test samples: the RC system is evaluated on the ability to retrieve input information memorized in the network states and to generalize to different values of experimental noise rather than to unseen samples. Specifically, the training and the test sets may share samples with the same nominal spike timing or rate. Since the same nominal perturbation is repeated, these samples differ only because of experimental noise. In practical applications, this evaluates the ability to infer properties of a known set of inputs.
    \item \textit{Unseen} test samples: the RC system is evaluated on its ability to generalize to unknown. Specifically, the test samples correspond to spike timings or rates that were unseen during the training procedure.
\end{enumerate}
For the \textit{classification} problem, we evaluate three variations:
\begin{enumerate}
    \item \textit{Single configuration} for all thresholds: all four binary classifications with different threshold are tackled by a single configuration of input wavelength and power at a time. The average classification accuracy obtained evaluates the ability of a single network input configuration to perform well for all four thresholds.
    \item \textit{1 feature per port}: this is a special case of the previous case, where the readout classifier is applied on only one value (or feature) per output port. Each one of the six features is obtained by averaging the readout over time, thus eliminating the time dependency in the used readout samples. Clearly, this variation is more challenging than the previous one, since the readout classifier has less available information. However, in practice, this variation relaxes the requirement on the speed of the photodetector and oscilloscope at the network output (less than \SI{1}{\mega\hertz} bandwidth is required), as the readout interval does not have to be temporally resolved.
    \item \textit{Multi-configuration}: for each threshold, we consider the best input configuration and then average the obtained accuracies. By allowing different input configurations to solve the binary classification, we can evaluate the best possible average accuracy achievable with the measured configurations.
\end{enumerate}

\subsection{ML results}
\label{subsec:MLperf}
\begin{figure*}[t!]
	\centering
	\includegraphics[width=0.8\textwidth]{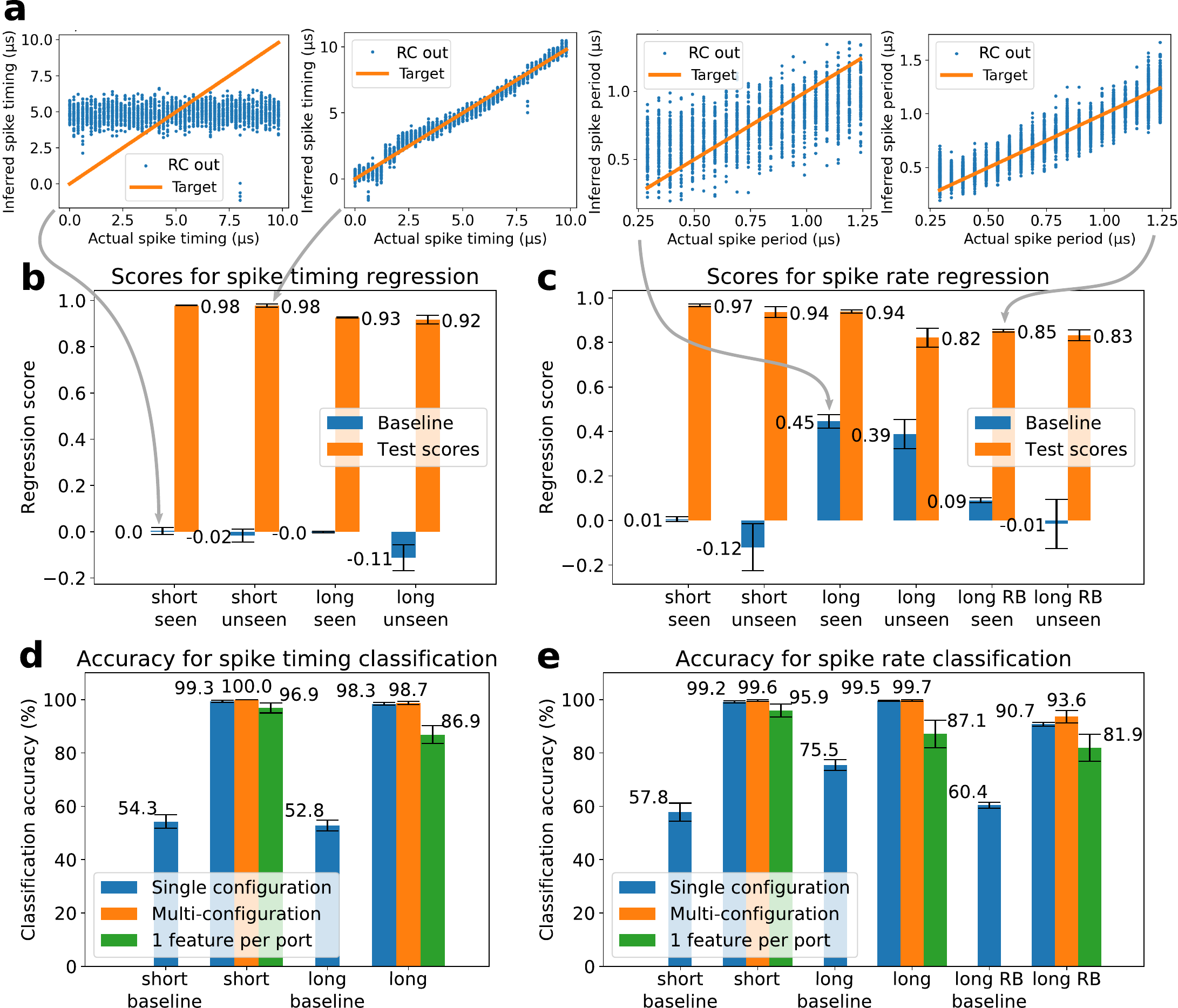}
	\caption{ \textbf{ML performances.} \textbf{a} Examples of the linear regression test prediction (light-blue dots) versus the actual ML target (orange lines). These examples correspond to the score values displayed in the bar plots of b or c, as indicated by gray arrows. \textbf{b, c, d, e} Bar plots of the best performances obtained by selecting the best input wavelength and power combinations, for each ML task and corresponding variations as described in  Section \ref{subsec:MLtasks} and used here to label each set of columns. The best input parameters, which best solve each ML task, are indicated by various symbols in the \SP maps of Fig. \ref{fig:2_SPmap}. In b,c, the orange columns refer to test score while the blue columns to the baseline. In c,d, the blue, orange and green columns refer respectivley to the single configuration, the multi-configuration and the 1 feature per port task variations as discussed in Section \ref{subsec:MLtasks}.
     The error bars in the regression tasks represent twice the standard deviation of the accuracies obtained for the different cross-validation folds, i.e. different ways of splitting the data into training and test datasets. The error bars in the classification tasks represent twice the standard deviation of the accuracy estimates for the 4 two-class binary problems, each requiring sample separation by a different threshold value.}
	\label{fig:4_results}
\end{figure*}

We present in Fig. \ref{fig:4_results} the best performance of our RC system on the different ML tasks. Optimum configurations (input wavelength and power) are shown by various symbols in the \SP maps of Fig. \ref{fig:2_SPmap}. To evaluate the performance of linear regression, we provide the \textit{coefficient of determination} of the prediction (referred to as \textit{regression score} in the plots), defined as one minus the ratio of the residual sum of squares to the total sum of squares (as implemented in the employed \textit{Ridge} function of the \textit{Scikit-learn} Python library \cite{pedregosa2011scikit}). In order to provide some intuitive insight into this performance measure, we first show some examples of the predictions of the linear regression test versus the actual ML targets (Fig. \ref{fig:4_results} a). These examples are matched by gray arrows to their score values in Fig. \ref{fig:4_results} b and c. 

Our RC system could perform \textit{spike timing regression} quite well in the short timescale case (Fig. \ref{fig:4_results} b). Importantly, the corresponding \textit{baseline} score (obtained from the aforementioned out-of-resonance measurement) is close to zero. This indicates that the obtained high score is solely due to the ability of our photonic reservoir to store and represent the input information through its multistability or \SP dynamics. In the longer timescale case, although still satisfyingly high, the regression score is reduced. Nevertheless, from this result we conclude that the spike timing information was stored in the photonic network state for at least \SI{75}{\micro\second}. 

The \textit{spike rate regression} is also performed well in the short timescale case (Fig. \ref{fig:4_results} c). We observe only a slight decrease in performance when the RC system was asked to generalize over unseen spike rates. In the long timescale case, even if the score is rather high for the \textit{seen} test samples case, the baseline score reaches 0.45, which is significantly larger than zero. Indeed, from the corresponding inference plot at the top, we can see that the linear readout can extract some information about the spike rate from the input signal (i. e. without the reservoir). This is due to the non-ideality of the optical modulator employed to generate the input waveforms. Indeed, in Fig. \ref{fig:3_tasks} d, it is possible to see that the pulse train perturbs the following NL stage power level. This introduces an unwanted memory effect that is exploited by the linear regression. Moreover, this relatively high baseline score also tells us that, if a simple memory effect such as the one introduced by the modulator can provide some non-negligible performance improvement, it might be that this task is not very challenging in computational terms. Indeed, considering that the perturbation with the lowest rate is more than four times longer than the one with the highest rate, the regression problem can be solved by directly looking at when the pulse train ends rather than at the actual spike rate. In order to remove this alternative and supposedly easier way to solve the regression, we have considered a more complex version of the problem, where the perturbation begins at random times between \SI{10}{\micro\second} and \SI{30}{\micro\second} w.r.t. the NL stage start (referred to as \textit{long timescale with random perturbation start} in Section \ref{subsec:MLtasks} and as \textit{long RB} in the x-axis labels of the bar plots in Fig. \ref{fig:4_results} c and e). In this case, the regression score is not very high (0.85 and 0.83 for the seen and unseen test samples cases, respectively). However, we successfully lower the baseline score to values close enough to the zero. Therefore, we conclude that the readout can extract the spike rate information from the reservoir state after at least \SI{40}{\micro\second}, even if with reduced performance. Regarding the sensitivity of coupled \MRs to spike rates, it is important to note that we have previously demonstrated non-fading \SP dynamics switching due to low spike rates in a system of three coupled \MRs, independent of the pulse train duration, in our earlier work \cite{biasi2024exploring}. 

The \textit{spike timing classification} achieves accuracies close to 100\% on both short and long timescales. These results significantly outperform the baseline accuracies of approximately 50 \%, which is the expected value for random guessing in binary classification. Consequently, we conclude that our RC system is effective in distinguishing early from late perturbations relative to a given time threshold by analyzing the reservoir response at least \SI{75}{\micro\second} after the perturbation ends. Notably, in the short timescale scenario, a high classification accuracy of 96.9 \% is achieved by reading only one value per output port, i.e., by averaging the network response over the readout time interval. This result is likely due in part to some fading memory that is not present in the long timescale scenario, where a significantly lower accuracy is observed.

The \textit{spike rate classification} also exhibits high accuracy in the short timescale case and a high baseline classification accuracy (75.5\%). Again, by imposing a random start time to the perturbing pulse trains, we lower the baseline classification accuracy to an acceptable level (60.4\%) and, concurrently, we improve the ability of the RC system to classify actual spike rates, rather than perturbation timings. This task is carried out with a satisfying best average accuracy of 93.6\%, by reading out the reservoir response after at least \SI{40}{\micro\second} since the NL stage beginning.

\section{Conclusion}
In this study, we present a first experimental demonstration of physical reservoir computing with all-optical non-fading memory, utilizing a passive network of silicon microring resonators as the photonic reservoir. This reservoir retains and processes information about its input excitation for an extended duration of at least several tens of microseconds. This capability is due to the sensitive yet stable dynamics and multistability of the 64 coupled silicon microresonators, which can be perturbed by an input signal in a non-fading manner. Crucially, this long-term memory effect can be leveraged to align the effective timescale of our physical neural network with that of the input signal, addressing a key challenge in neuromorphic computing and in reservoir computing. Additionally, our integrated photonic reservoir offers several benefits: a compact size (0.15 mm$^2$), relatively low power consumption (ranging from a few to tens mW), reconfigurability by altering the input’s excitation wavelength or power without requiring external wiring, ease of fabrication, and multi-wavelength operations that enable network expansion and/or parallel processing.

We demonstrated that the proposed reservoir computing system can handle both regression and classification tasks related to the timing of an input single pulse and the spike rate of an input spike train, across two different timescales (with approximately a fivefold difference). The first task is particularly relevant in distributed optical fiber sensing, where different optical pulse timings correspond to different perturbation locations along the fiber sensor. Additionally, encoding information into spike timing or spike rate are two primary methods for operating spiking neural networks. Therefore, our dynamic photonic network shows potential as an all-optical interface for reconfigurable and energy-efficient decoding or processing of photonic spiking processors or sensors. Overall, our work presents a practical and versatile new approach to extending memory in photonic reservoir computing, making it applicable to sensor signals with much slower timescales.

\section{Methods}
\label{sec:methods}

\subsection{Experimental setup and photonic integrated circuit}
\label{subsec:methods_setup}

\begin{figure*}[t!]
	\centering
	\includegraphics[width=0.7\textwidth]{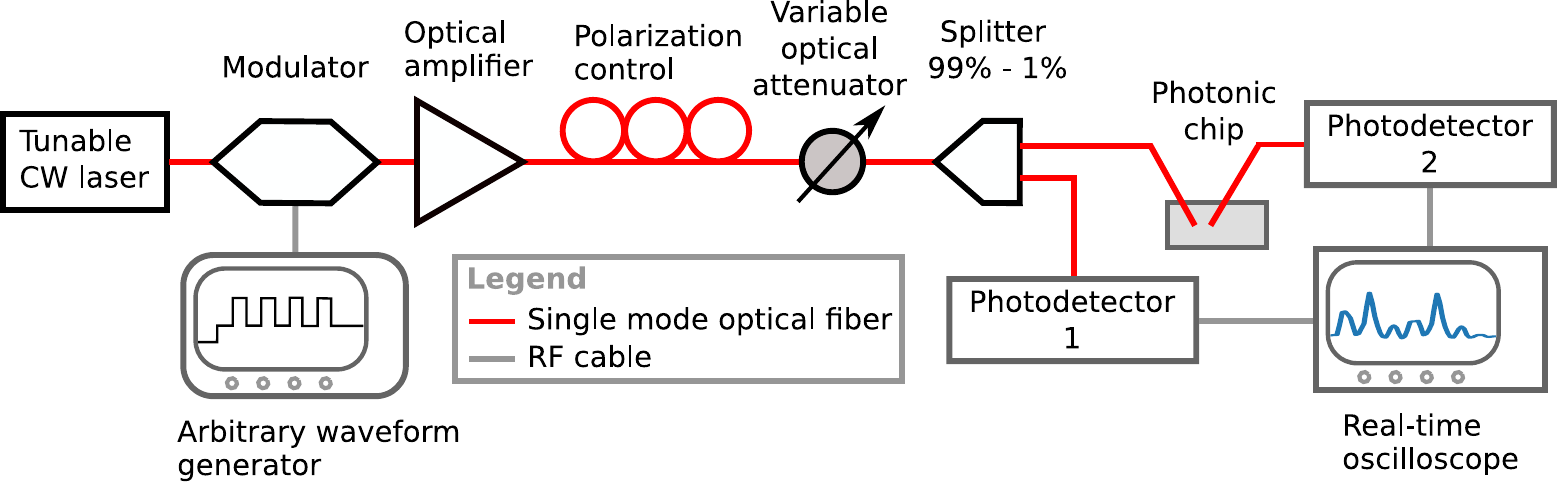}
	\caption{\textbf{Experimental setup.} Laser light (wavelength around \SI{1550}{nm}) is modulated into the desired waveform and injected into the on-chip photonic network. The input wavelength is selected by using a tunable continuous wave laser, while the input power is set via a variable optical attenuator. Before the photonic chip, we placed a 99\% - 1\% splitter, so that we could read out the 1\% port with photodetector 1, allowing us to monitor the average input power. Finally, the photonic network output is read out by photodetector 2. Red lines refer to the optical signal path, while the gray lines to the Rf signal path.}
 \label{fig:5_setup}
 \end{figure*}

We used a standard photonic setup to inject and collect an optical signal via a photonic integrated circuit (PIC) with a bandwidth of approximately \SI{500}{\mega\hertz} (see schematics in Fig. \ref{fig:5_setup}). The input optical signal was generated by a fiber-coupled continuous-wave tunable laser (Pure Photonics) operating at the C-band and modulated by an electro-optic modulator (EOM, iXblue model MXAN-LN-10) driven by an arbitrary waveform generator (AWG, Spectrum model DN2.663-02). The amplitude-modulated optical signal was then amplified by an erbium-doped fiber amplifier (EDFA, Thorlabs) and attenuated to the desired power level via an electronic variable optical attenuator (VOA). A polarization control stage fixed the proper (TE, transverse electric) polarization.

Then, a 1\% fiber tap coupled to a slow photodetector (PD, 30 kHz bandwidth, New Focus model M2033) monitored the average input optical power. The other 99\% signal was injected into the PIC through a cleaved fiber placed on a three-axis linear piezoelectric stage. The PIC transmission was collected by another cleaved optical fiber, interfaced through another VOA, to a fast PD (\SI{600}{\mega\hertz}, Menlosystem model FPD610-FC-NIR). This second VOA protected the fast PD from exceedingly high optical power. The electrical signals from the two PDs were read by an oscilloscope (Picoscope model 6000) and stored in a computer, which also controlled the instrumentation. Note that the chip temperature was stabilized by a thermostat system, whose temperature was controlled by a proportional-integral-derivative controller connected to a Peltier cell and a 10 k$\Omega$ thermistor.

The PIC is composed by a \MRs array based on silicon-on-insulator waveguides with a silicon core cross-section of \SI{450}{\nano\meter} \(\times\) \SI{220}{\nano\meter}, embedded in a silica cladding. The \MRs have a racetrack shape with a bend radius of 7 µm and straight coupling sections of \SI{0.71}{\micro\meter}. The gap between the bus waveguides and the \MR waveguide is \SI{0.2}{\micro\meter} long. The distance between the centers of adjacent \MRs on the same line is \SI{22.7}{\micro\meter}, while \MRs on adjacent lines are horizontally displaced by \SI{11.35}{\micro\meter} (see Fig. \ref{fig:1_network}). The PIC was fabricated by IMEC (Leuven, Belgium).

\subsection{Machine learning aspects}
\label{subsec:methods_ML}
For the single pulse timing inference ML task (with reference to Section \ref{subsec:MLtasks}), we measured the response of the photonic reservoir to 5 or 6 repetitions (depending on how many fit our acquisition time window) of the input signal. The input signal was composed by 10 repetitions of a randomly ordered sequence of 50 NL stages, each with different perturbation pulse timings. Thus, for each measurement and ML task, we acquired and processed 2500 or 3000 samples. Here, we stress the importance of randomizing the order of samples with respect to the label values to avoid shortcut learning due to correlation between labels and drifts in experimental conditions \cite{geirhos2020shortcut}. We employed a similar input  scheme for the spike rate inference (in this case, 20 different spike rates, i. e. 1000 or 1200 samples). 

The spike rate inference ML task with random perturbation beginning is intrinsically more complex than the others. This is due to the use of randomness in the beginning time as an additional degree of freedom. Therefore, in this case, we needed a larger number of ML samples and thus we acquired the network response to 5 or 6 repetitions of the following input signal: 50 repetitions of a randomly ordered sequence of 20 NL stages, each with a different pulse rate and start of the perturbation pulse train. We thus obtained 5000 or 6000 samples. It should be added that we excluded from the ML processing the first 5 samples (or NL stages) at the start of each input repetition to avoid signal distortions by the non-idealities in the modulator after long pauses.

The training of the linear readout in the RC system was done with the \textit{Scikit-learn} Python library \cite{pedregosa2011scikit}, specifically the \textit{Ridge} and \textit{Logistic Regression} functions for linear regression and linear classification, respectively.  Moreover, we used the following pipeline:
\begin{enumerate}
    \item Downsample the readout waveforms, which are the network responses recorded during the readout intervals. We used the signal.decimate function from the SciPy Python library with ratios of 400, 40, 20, and 10 for the short timescale cases, and 1600, 160, 80, and 20 for the long timescale cases, respectively. Consequently, for each measurement, we obtained four different datasets, each corresponding to a different downsampling ratio.
    \item For each downsampling ratio value, initiate the outer 5-fold cross-validation loop, where the training and test datasets are separated. We also standardize the features using the feature averages and standard deviations calculated from the training set. 
    \item Select the best L2 normalization strength in the inner cross-validation loop from 8 values, ranging from \(10^{-8}\) to \(10^{-1}\), using factor 10 steps.
    \item With optimized regularization strength, train and test the ML model, closing the outer cross-validation loop.
\end{enumerate}
It is important to emphasize that for each ML task, we repeated this ML pipeline for every combination of downsampling ratio, input wavelength, and input power. The test scores presented in Fig. \ref{fig:4_results} represent the best results selected from this parameter space.

\section{Data availability}
The data supporting the results of this study are available from the corresponding author upon reasonable request.

\section{Code availability}
The Python code employed to generate the input waveforms and to process the output waveforms will be uploaded to the public repository Zenodo, once this work is published. In the meantime, this code could be provided by the corresponding author upon reasonable request.

\bibliographystyle{unsrt}
\bibliography{references}  %%% Uncomment this line and comment out the ``thebibliography'' section below to use the external .bib file (using bibtex) .





% \end{thebibliography}

\section{Acknowledgments}
This project has received funding from the European Research Council (ERC) under the European Union’s Horizon 2020 research and innovation programme (grant agreement No. 788793, BACKUP). A. Lugnan acknowledges funding by the European Union under GA n°101064322-ARIADNE. S. Biasi acknowledges the cofinancing of the European Union FSE-REACT-EU, PON Research and Innovation 2014–2020 DM1062/2021. A. Foradori acknowledges funding by the European Union under GA n°101070238-NEUROPULS.
Views and opinions expressed are however those of the author(s) only and do not necessarily reflect those of the European Union or The European Research Executive Agency. Neither the European Union nor the granting authority can be held responsible for them.

\section{Author contributions}
A.L. and S.B. conceived the experiment. A.L. and A.F. designed and performed the experimental measurements. S.B., A.F. and A.L. prepared the experimental setup.  A.L. designed the integrated photonic circuit under the supervision of P.B.. S.B. analyzed the frequency of the self-pulsing response traces and created the corresponding figure. A.L. designed and performed the machine learning analysis. A.L. wrote the manuscript. L.P. and P.B. supervised the work. All authors contributed to the revision of the manuscript.

\section{Competing interests}
The authors declare no competing interests.

%\section{\textcolor{red}{Funding}}
%\sbi{H2020 European Research Council (788793, BACKUP).}

\end{document}